\documentclass[twocolumn]{jpsj3}
\usepackage{graphicx,amsmath,amssymb,bm}
\usepackage{braket}

\allowdisplaybreaks 

\usepackage{color}
\definecolor{Green}{rgb}{0,0.7,0}

\newcommand{\ET}{$\alpha$-(BEDT-TTF)$_2$I$_3$}

\newcommand{\bk}{\bm{k}}

\newcommand{\bkD}{\bm{k}_{\rm D}}
\newcommand{\eD}{\epsilon_{\rm D}}
\newcommand{\ep}{\epsilon }

\newcommand{\g}{\gamma }
\newcommand{\bq}{\bm{q}}

\newcommand{\muz}{\mu_0}

\newcommand{\spcx}{$\sigma_x(\ep,T)$\; }
\newcommand{\spcy}{$\sigma_y(\ep,T)$\; }
\newcommand{\spcn}{$\sigma_\nu(\ep,T)\; $}
\newcommand{\spcym}{$\sigma_y(\ep,T)$}

\begin{document}

\title{
Seebeck Effect of Dirac Electrons in Organic Conductors under Hydrostatic Pressure Using a Tight-Binding Model Derived from First Principles
}

\author{
Yoshikazu Suzumura,
$^{1}$\thanks{E-mail: suzumura@s.phys.nagoya-u.ac.jp}
 Takao Tsumuraya,
$^{2}$\thanks{E-mail: tsumu@kumamoto-u.ac.jp}
and 
Masao \surname{Ogata}
$^{3,4}$\thanks{E-mail: ogata@phys.s.u-tokyo.ac.jp}
}
\inst{
$^1$Department of Physics, Nagoya University,  Nagoya 464-8602, Japan \\
$^2$Magnesium Research Center, Kumamoto University, Kumamoto 860-8555, Japan\\
$^{3}$Department of Physics, University of Tokyo, Bunkyo, Tokyo 113-0033, Japan \\
$^{4}$Trans-scale Quantum Science Institute, University of Tokyo, Bunkyo, Tokyo 113-0033, Japan
}

\recdate{January 5, 2024;  accepted February 26, 2024}

\abst{
 The Seebeck coefficient is examined for two-dimensional Dirac electrons 
 in the three-quarter filled  organic conductor \ET \ [BEDT-TTF denotes  
bis(ethylenedithio)tetrathiafulvalene] under hydrostatic pressure, 
  where the Seebeck  coefficient $S$ is  proportional to the ratio of the thermoelectric conductivity $L_{12}$ to the electrical conductivity $L_{11}$, i.e., $S=L_{12}/TL_{11}$ with $T$ being the temperature.  
 We present an improved tight-binding model in two dimensions with  transfer energies determined from  first-principles density functional theory calculations with  an  experimentally determined crystal structure. 
 The $T$ dependence of the Seebeck  coefficient is calculated by adding impurity  and electron--phonon  scatterings.   
  Noting a zero-gap state due to the Dirac cone, which results in a competition from contributions between the conduction and valence bands, we show positive $S_x$ and $S_y$ at finite temperatures  and  analyze them  in terms of  spectral conductivity. 
  The relevance of the calculated $S_{x}$  (perpendicular to the molecular stacking axis) to the experiment is discussed.
  }


\maketitle

\section{Introduction}
 Two-dimensional massless Dirac fermions with linear spectra from  Dirac points have been studied extensively.  
  Among them, a bulk material has been found in  an  organic conductor,~\cite{Kajita_JPSJ2014} \ET,  where BEDT-TTF denotes  
bis(ethylenedithio)tetrathiafulvalene.\cite{Mori1984}
 This material exhibits a zero-gap state (ZGS)\cite{Katayama2006_JPSJ75} and  the transport properties are understood from the density of states (DOS), which reduces linearly to zero at the Fermi energy.\cite{Kobayashi2004} 
 The explicit band structure of the Dirac cone was calculated using  a tight-binding (TB) model, where transfer energies under various  pressures are estimated  by the extended H\"uckel method.\cite{Kondo2005,Kondo2009} 
Furthermore, Dirac electrons in the organic conductor were examined using a two-band model.~\cite{Kobayashi2007,Goerbig2008}

A first-principles calculation based on density functional theory (DFT) confirmed the existence of the Dirac cone in~\ET~at ambient pressure.\cite{Kino2006}
Later, a DFT band structure was calculated~\cite{Alemany} from the crystal structure determined by X-ray diffraction measurements under a hydrostatic pressure of 1.76GPa.~\cite{Kondo2009} Despite these investigations, the impact of pressure on the transfer energies calculated through first-principles methods has not been explored. 

Characteristic properties  of  Dirac fermions appear in the temperature ($T$) dependences of various physical quantities. 
The $T$-linear behavior of  the magnetic susceptibility\cite{TakanoJPSJ2010,Hirata2016} and the sign change of the Hall coefficient,\cite{Tajima_Hall2012} have been well understood by theories using the four-band model\cite{Katayama_EPJ} and the two-band model\cite{Kobayashi2008}, respectively. 
However, an almost $T$-independent conductivity\cite{Kajita1992,Tajima2000,Tajima2002,Tajima2007,Liu2016} cannot be understood in a model with only impurity scattering.\cite{Neto2006} 
If we consider the effect of the electron--phonon (e--p) interaction, a nearly constant conductivity can be understood using a simple two-band model of the Dirac cone without tilting.\cite{Suzumura_PRB_2018}
Since the energy band in the actual organic conductor shows a deviation 
 from a linear spectrum,~\cite{Katayama2006_cond}  we have also examined the conductivity using the TB model with the  transfer energies obtained from  \ET.\cite{Suzumura_JPSJ2021}

In this paper, we examine theoretically the Seebeck coefficient of \ET, 
  which has been observed  experimentally under  hydrostatic  pressure,\cite{Tajima_Seebeck,Konoike2013} along the $x$-direction (perpendicular to the molecular stacking axis). 
The formula of Seebeck effects has been established  using a linear response theory.~\cite{Kubo_1957,Luttinger1964,Ogata_Fukuyama} In \ET, the Seebeck coefficient was studied  using an extended Hubbard model with transfer energies at ambient pressure and  Coulomb interactions.~\cite{Kobayashi_2020}
  However, the calculated Seebeck coefficient was not along the $x$-direction.
Recently, we have examined the Seebeck coefficient along the $x$-direction,~\cite{YS_Ogata_cond-mat} i.e.,  the same  direction as that in an experiment by taking a uniaxial pressure. 
Although qualitatively the same  $T$ dependence as that in the experiment  has been obtained,  the case of hydrostatic pressure remained as a problem. 
 A model calculation using the transfer energies  obtained   by the extended H\"uckel method~\cite{Kondo2009} and by a previous first-principles calculation at ambient pressure\cite{Kino2006}  gives  a negative  Seebeck coefficient,~\cite{YS_Ogata_cond-mat}  whereas experimental results show  a positive Seebeck coefficient at finite temperatures.\cite{Tajima_Seebeck,Konoike2013}
Therefore, in this paper, we perform  first-principles DFT calculations 
using the experimentally obtained crystal structure at 1.76 GPa\cite{Kondo2009}
and derive  an $ab$ $initio$ tight-binding model. Then, we reexamine the Seebeck coefficient using  this improved  model.
 
In Sect. 2,  the model  and the energy band of the ZGS  in \ET\ are given.  
  The $T$ dependence of the chemical potential is shown by calculating the DOS.
  In Sect. 3,  the  $T$ dependence of the Seebeck coefficients for 
  both the  $x$- and $y$-directions is demonstrated  and  analyzed in terms of   spectral conductivity. The electric conductivity is also shown.  In Sect. 4, a summary is given and the  present calculations  are  compared  
 with  experimental results. 

\section{Model and Electronic States}
\subsection{Model Hamiltonian} 
We consider a two-dimensional Dirac electron system per spin,
 which is given by 
\begin{equation}
H = H_0  + H_{\rm p} +  H_{\rm e-p} +H_{\rm imp} \; . 
\label{eq:H}
\end{equation}
Here, $H_0$ describes a TB model of the  organic conductor  consisting of four  molecules per unit cell. The second term denotes the harmonic phonon given by  
 $H_{\rm p}= \sum_{\bq} \omega_{\bq} b_{\bq}^{\dagger} b_{\bq}$  with $\omega_{\bq} = v_s |\bq|$ and  $\hbar$ =1. The third term is  the e--p interaction with a coupling constant $g_{\bq}$,
\begin{equation}
 H_{\rm e-p} = \sum_{\bk, \g} \sum_{\bq}
   g_{\bq} c_\g(\bk + \bq)^\dagger c_{\g}(\bk) 
(b_{\bq} + b_{-\bq}^{\dagger})
\; ,
\label{eq:H_e--p}
\end{equation}
 with $c_{\g}(\bk) = \sum_{\alpha} d_{\alpha \g}a_{\alpha}(\bk)$, which is obtained by the diagonalization of $H_0$ as shown later. The e--p scattering is considered  within  the same band (i.e., intraband) owing  to the energy conservation with $v \gg v_s$,~\cite{Suzumura_PRB_2018} where $v$ denotes the average velocity of the Dirac cone. The last term of Eq.~(\ref{eq:H}), $H_{\rm imp}$, denotes a normal  impurity  scattering.
\begin{figure}
  \centering
\includegraphics[width=6cm]{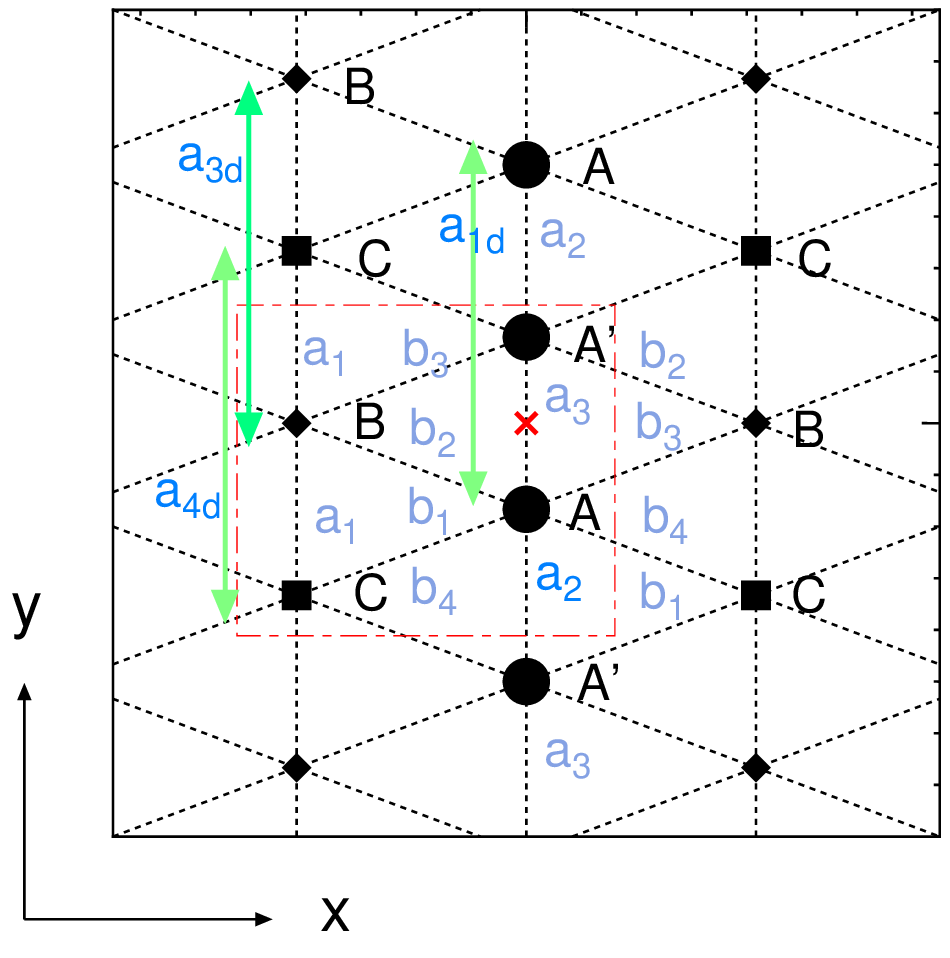}
     \caption{(Color online)
Crystal structure of \ET. There are four molecules A, A', B, and C in a unit cell (dot-dashed line). The unit cell forms a square lattice, whereas  each molecule is located on the triangular lattice. Transfer energies are given as $a_1, \cdots, b_4$  for the nearest-neighbor (NN) sites and $a_{1d}$, $a_{3d}$, and $a_{4d}$ for the next-nearest-neighbor (NNN) sites. The cross denotes an inversion center between A and A'. $x$ ($y$) denotes a coordinate  perpendicular to (along) a molecular stacking  direction. 
 }
\label{fig1}
\end{figure}

The TB model $H_0$ without spin degrees of freedom  is expressed as 
\begin{eqnarray}
H_0 &=& \sum_{i,j = 1}^N \sum_{\alpha, \beta}
 t_{i,j; \alpha,\beta} a^{\dagger}_{i,\alpha} a_{j, \beta},
\label{eq:Hij}
\end{eqnarray}
where $t_{i,j; \alpha,\beta}$ depicts  the taransfer energy. 
$N$ is the total number of unit cells  and   $a^{\dagger}_{i, \alpha}$ denotes a creation operator of an electron  on a molecule $\alpha$ in the $i$-th unit cell.  Figure \ref{fig1} shows a crystal structure of \ET\ used in the TB model. In the unit cell,  there are four BEDT-TTF molecules  $\alpha$  [= A(1), A'(2), B(3), and C(4)]. 

\subsection{Transfer energies of TB model obtained by DFT}
We derive the quantity $t_{i,j; \alpha,\beta}$ in Eq.~(\ref{eq:Hij}) from first-principles DFT calculations. The explicit expression for the transfer energy is provided by
\begin{eqnarray}
t_{\alpha,\beta}(\mathbf{R})=\langle\phi_{\alpha,0}|H_{k}|\phi_{\beta,\mathbf{R}} \rangle ,
\label{equ1}
\end{eqnarray}
where $H_{k}$ is the one-body part of the \textit{ab initio} Hamiltonian for \ET.~\cite{Kino2006} $\phi_{\alpha,\mathbf{R}}$ denotes the maximally localized Wannier function (MLWF) spread over the BEDT-TTF molecule with $\alpha$ ($\beta$) representing distinct MLWFs and $\mathbf{R}$ indicating the location of the $j$-th unit cell relative to the $i$-th unit cell~\cite{Tsumuraya_Suzumura}. 
The crystal structure employed in our calculations is based on an experimental structure measured at 1.76 GPa~\cite{Kondo2009}, with structural optimization performed for hydrogen positions. 

Equation (\ref{equ1}) reveals that $t_{i,j; \alpha,\beta}$ relies  not on the respective site but solely on the difference between the $i$-th and $j$-th sites. Subsequent to acquiring Bloch functions through the DFT calculations,\cite{HK1964, KS1965} the MLWFs were constructed utilizing the \texttt{wannier90} code~\cite{Marzari1997, Isouza2001}. Transfer energies were then computed on the basis of the overlaps between the four MLWFs. It is noteworthy that the center of each MLWF is positioned at the midpoint of the central C = C bonds in each BEDT-TTF molecule.
The present DFT calculations are based on a pseudopotential technique and plane wave basis sets via the projected augmented plane wave method,\cite{PAW1994} as implemented in the Vienna \textit{ab initio} simulation package (VASP).\cite{Kresse_VASP1996, Kresse_Joubert_PAW1999} 
The exchange-correlation functional used in this work is the generalized gradient approximation (GGA) proposed by Perdew, Burke, and Ernzerhof (PBE).\cite{GGAPBE}
We set the cutoff energies to 400 and 645 eV for plane waves and augmentation charge, respectively. We employed a 4 x 4 x 2 uniform $\bk$-point mesh with a Gaussian smearing method during self-consistent loops.

The obtained transfer energies  $a_1, \cdots, b_4$, $a_{1d}, a_{3d}$, and $a_{4d}$ in Fig.~\ref{fig1}  are listed  in Table \ref{table_1}.  The quantities 
 $t_{\rm AA}$, $t_{\rm A'A'}$, $t_{\rm BB}$, and $t_{\rm CC}$ denote the diagonal elements  corresponding to the on-site potential energy. 
 Since the origin of  energy is arbitrary,  we  take only  the   on-site potentials $V_{\rm B}$ and $V_{\rm C}$  measured from that of  A and  A', which are defined by 
$V_{\rm B} \equiv t_{\rm BB} - t_{\rm AA}$ and  $V_{\rm C} \equiv t_{\rm CC} - t_{\rm AA}$, with $t_{\rm AA} = t_{\rm A'A'}$ .
The unit of energy is taken as eV.

\begin{table}[b]
\centering
\caption{First-principles TB model parameters in eV, which are obtained 
from  Eq.~(\ref{equ1}).}
\begin{minipage}{0.5\hsize}
\centering
\label{Transfer_alpha}
\begin{tabular}{crrrrrrr}
\hline\hline
$t_{i,j; \alpha,\beta} (i \not=j)$  &   \\ 
\hline
$a1$  & $- 0.0122$    \\
$a2$  & $- 0.0239$   \\ 
$a3$  & 0.0520   \\ 
\hline        
$b1$  & 0.1508   \\ 
$b2$  & 0.1335    \\ 
$b3$  & 0.0568    \\ 
$b4$  &  0.0194   \\ 
\hline        
$a_{1d}$ & 0.0130   \\
$a_{3d}$ & 0.0032    \\ 
$a_{4d}$ & 0.0185    \\ 
\hline\hline
\\
\end{tabular}
\end{minipage}
\begin{minipage}{0.45\hsize}
\centering
\begin{tabular}{cr}
\hline
\hline
$ t_{i,i; \alpha,\alpha}$ & \\ 
\hline
$t_{\rm AA}$  & 4.744677 \\ 
$t_{\rm A'A'}$  & 4.744677 \\ 
$t_{\rm BB}$  & 4.693977 \\
$t_{\rm CC}$  &  4.678874 \\
\hline
\hline
\label{table_1}
\end{tabular}
\end{minipage}
\end{table}

\subsection{Electronic states}
 $H_0$  is diagonalized by  $H_0 |\g> =  E_\g |\g> $, i.e.,  
\begin{eqnarray}
\label{eq:eigen_eq}
\sum_{\beta} h_{\alpha,\beta}(\bk) d_{\beta \g(\bk)}
   &=& E_{\g}(\bk) d_{\alpha \g} (\bk)  \; , 
 \label{eq:E_alpha}
\end{eqnarray}
where  $h_{\alpha,\beta}(\bk)$ is the Fourier transform of Eq.~(\ref{eq:Hij}) 
  given in  Appendix A. $E_\gamma(\bk)$  $(\gamma = 1, \cdots, 4)$  denotes the energy in the descending order. The Dirac point ($\bkD$)  is  obtained from  
\begin{eqnarray}
\label{eq:ZGS}
E_1(\bkD) = E_2(\bkD)= \eD \; ,
\end{eqnarray}
 where $E_1(\bk)$ denotes the  conduction band  and $E_2(\bk)$ denotes the  valence band.  The ZGS is found  when  $\eD$ becomes equal to  the chemical potential $\mu$.  From the three-quarter-filled condition,  $\mu$ is calculated by 
\begin{eqnarray}
  \frac{1}{N} \sum_{\bk} \sum_{\gamma}  f(E_{\gamma}(\bk))=  3 \; ,  
  \label{eq:mu}
\end{eqnarray}
where  $f(\ep)= 1/(\exp[(\ep -\mu)/k_{\rm B}T]+1)$ with $T$ being temperature 
 and the Boltzmann constant taken as $k_{\rm B }= 1$. 

\begin{figure}
  \centering
\includegraphics[width=4cm]{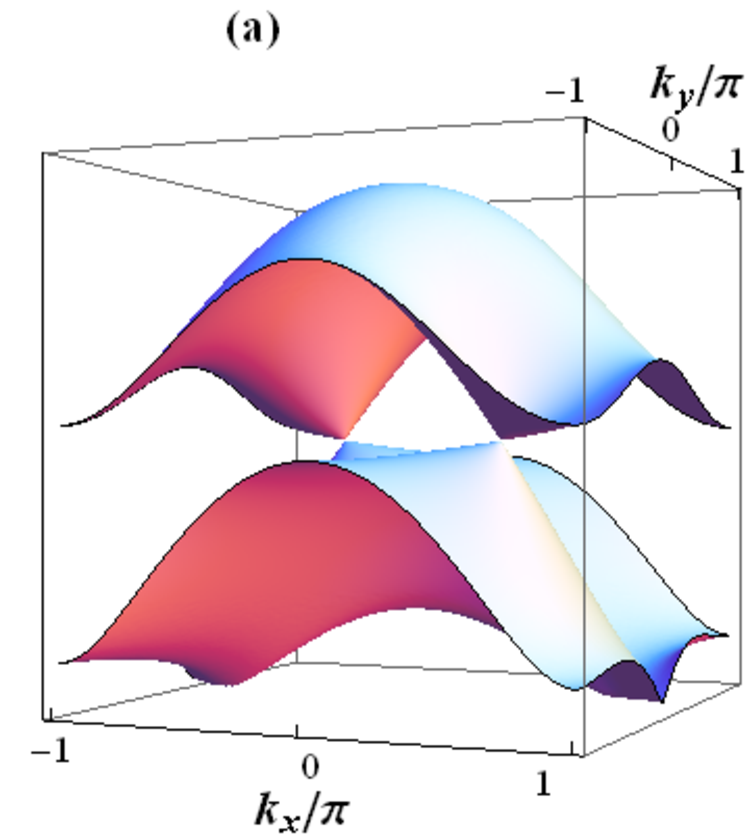}
\includegraphics[width=4cm]{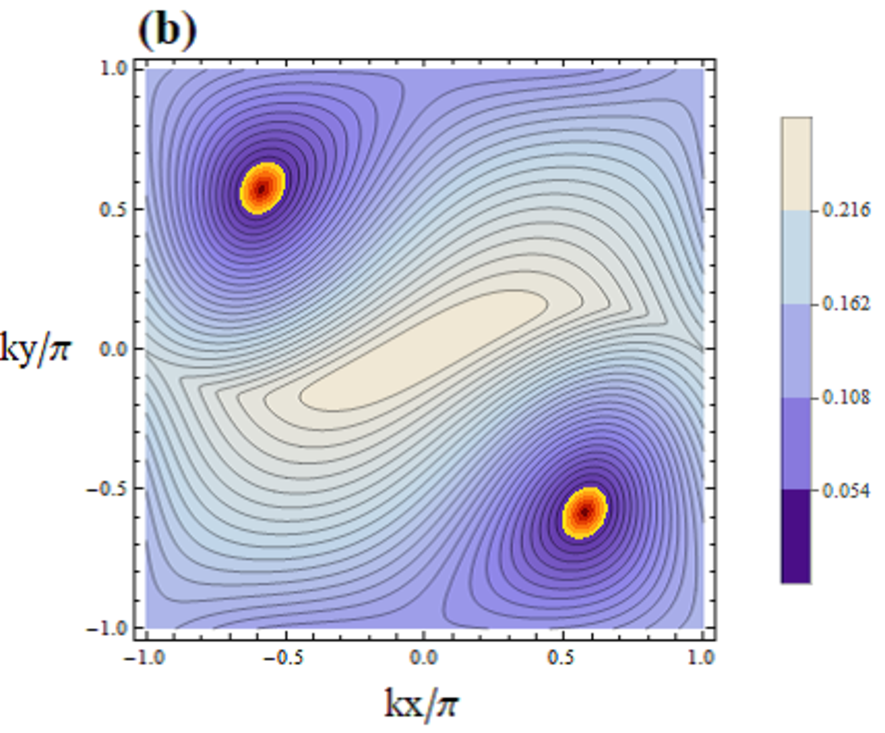}\\
\includegraphics[width=4cm]{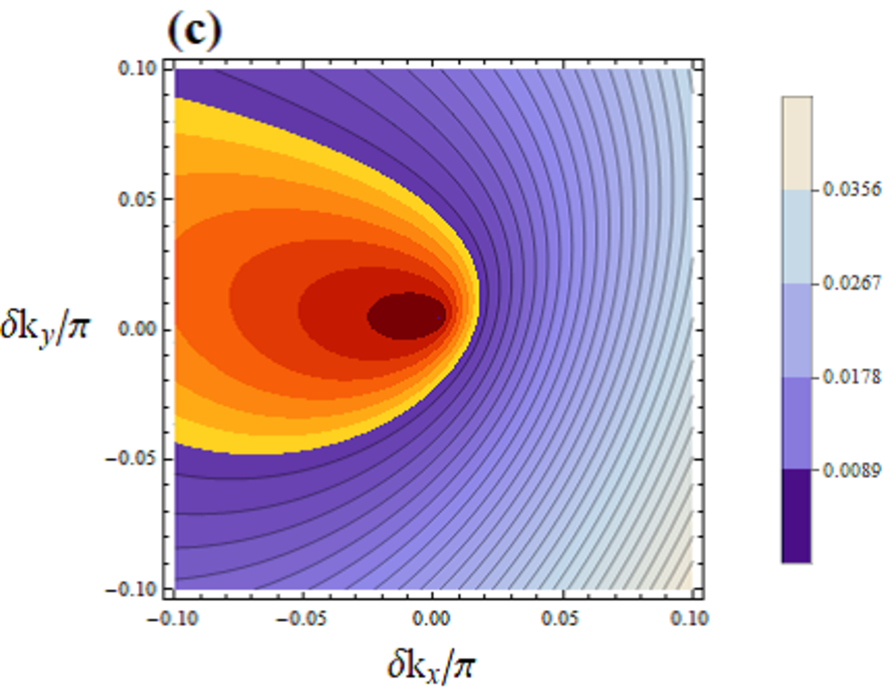}
\includegraphics[width=4cm]{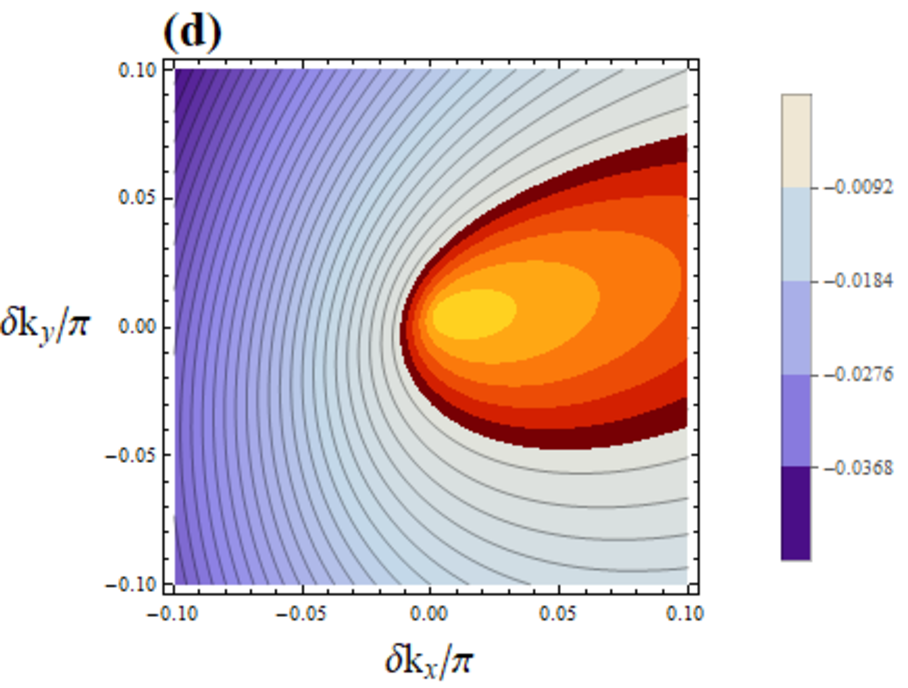}\\
\includegraphics[width=4cm]{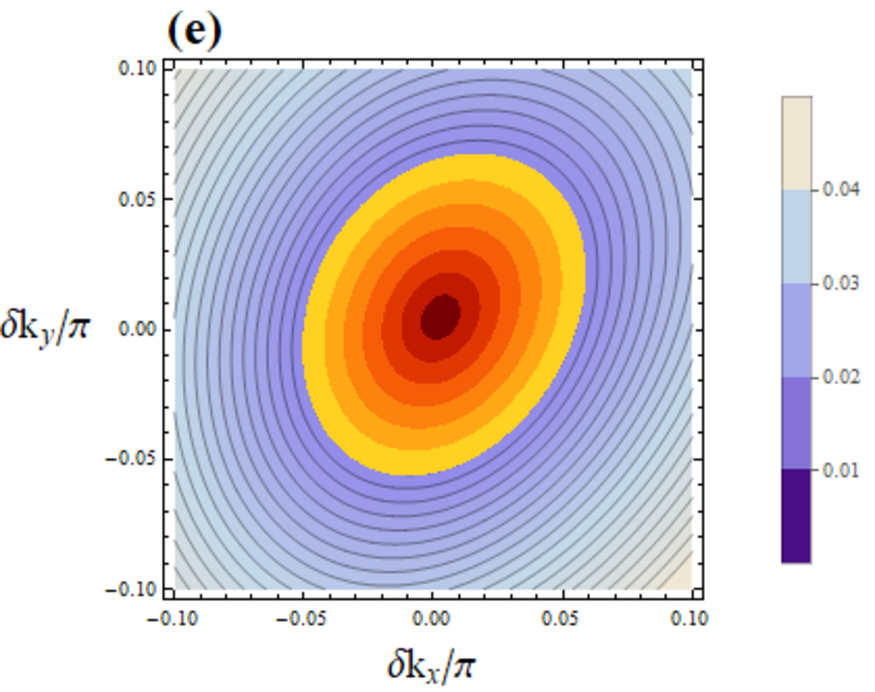}
     \caption{(Color online)
(a) Conduction and valence bands given by $E_1(\bk)$ (upper band) and $E_2(\bk)$ (lower band). 
(b) Contour plots of $E_1(\bk) - E_2(\bk)$, where the yellow region including the Dirac point $\bkD$ is given by $0< E_1(\bk) - E_2(\bk) <  0.03$. 
 (c) Contour plots  of  $ E_1(\bk)$ around $\bkD$, where $\delta \bk = \bk -\bkD$. The yellow region corresponds  to   $ 0 <   E_1(\bk) - \eD <  0.005$. 
(d) Contour plots of   $ E_2(\bk)$ around $\bkD$.  The  yellow region corresponds  to   $ -0.005 <   E_2(\bk) - \eD <  0$.
(e)  Contour plots of $ E_1(\bk) - E_2(\bk)$ around $\bkD$. The yellow region corresponds  to   $ 0  <   E_1(\bk) - E_2(\bk) <  0.02$.
    }
\label{fig2}
\end{figure}
  Using   the TB model with transfer energies given in Table \ref{table_1},  we examine the energy bands  obtained from Eq.~(\ref{eq:E_alpha}).
 The site potentials measured from that at A and A' are given as  $V_B = -0.0507$ and  $V_C = -0.0650$. In the following, two bands of  $E_1(\bk)$ and $E_2(\bk)$ are examined since  $E_3(\bk)$ and $E_4(\bk)$ are located far below  the chemical potential. The calculated  energy bands are shown in Figs.~\ref{fig2}(a)--\ref{fig2}(e). 

Figure \ref{fig2}(a) shows $E_1(\bk)$ (upper band) and $E_2(\bk)$ (lower band) for \ET, which contact at the Dirac points $\pm \bkD = \pm (-0.58, 0.57)\pi$  with the energy $\eD$ forming the ZGS.  Figure \ref{fig2}(b) shows contour plots of  $E_1(\bk) - E_2(\bk)$ providing $\pm \bkD$.\cite{Piechon2015} Figures \ref{fig2}(c) and \ref{fig2}(d) show contour plots  of the conduction band  $E_1(\bk)$ and the valence band $E_2(\bk)$, respectively,  with $\delta \bk = \bk -\bkD$. 
 Figure \ref{fig2}(e) shows magnified contour plots of $ E_1(\bk) - E_2(\bk)$ around $\bkD$. The ellipsoid of the contour suggests an anisotropy of the velocity of the Dirac cone  with $v_x > v_y$. 

\begin{figure}
  \centering
\includegraphics[width=6cm]{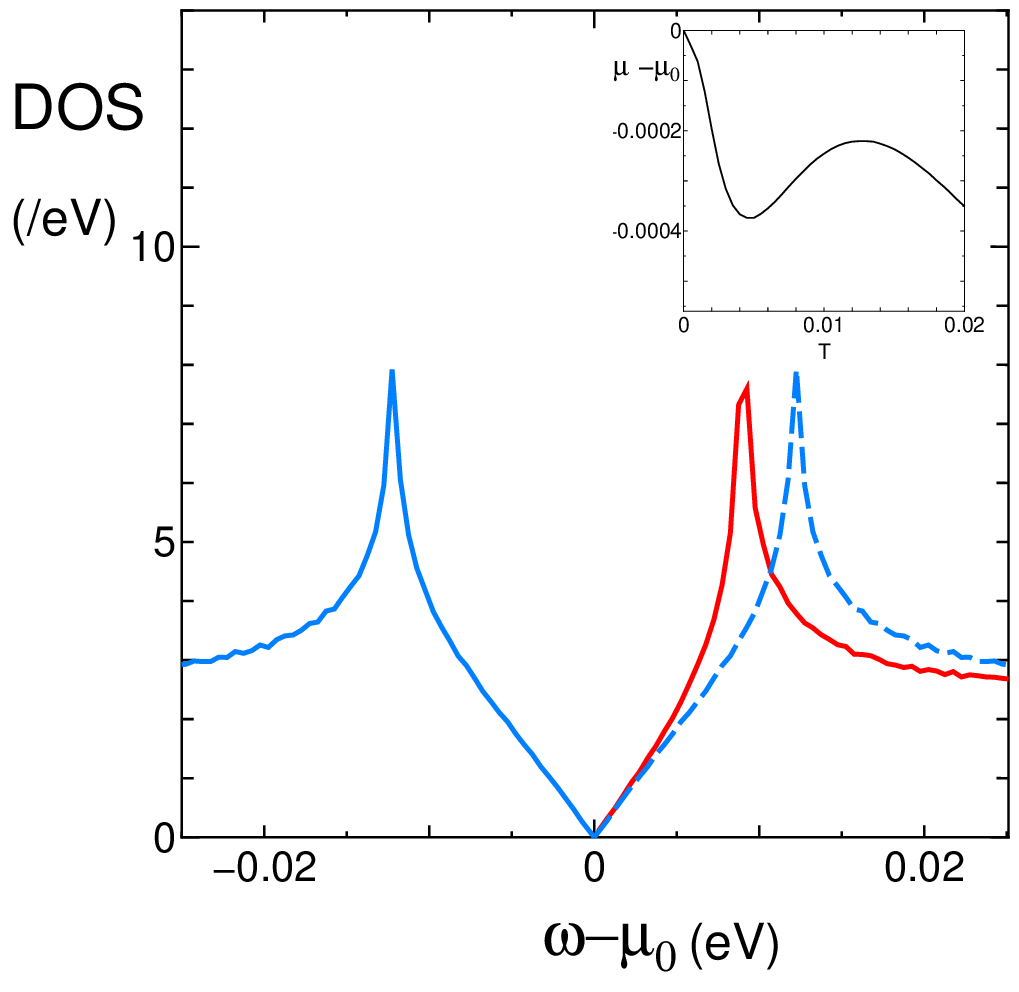}
     \caption{(Color online )
 DOS as a function of $\omega - \muz$, where $\muz= \eD = 0.1679$ is  $\mu$ at $T = 0$.  The blue dashed line denotes the symmetrized DOS using $D(\omega)$ with $\omega<\muz$. 
 For the fixed  $|\omega-\muz| (< 0.01)$, DOS with $\omega>\muz$ (red solid line)  is larger than that with  $\omega<\muz$ (blue line). 
    The inset denotes $\mu -\muz$ as a function of $T$.}
 \label{fig3}
\end{figure}

 Using $E_{\gamma}(\bk)$, we calculate DOS as 
\begin{eqnarray}
  D(\omega) = \frac{1}{N} \sum_{\bk} \sum_{\gamma}  
 \delta (\omega - E_{\gamma}(\bk)),
  \label{eq:DOS}
\end{eqnarray}
which is shown in Fig.~\ref{fig3} as a function of $\omega - \muz$ with  $\muz$ being $\mu$ at $T = 0$. There is an asymmetry of DOS around $\omega = \muz$; DOS for the conduction band ($\omega > \muz$) is slightly larger than that for the valence band ($\omega < \muz$). The inset shows the $T$ dependence of the chemical potential $\mu$. Owing to the asymmetry of DOS, $\mu$ becomes smaller than $\muz$ at finite temperatures.

\section{Seebeck Effects}
\subsection{Seebeck coefficient}
Using the  linear response theory,~\cite{Kubo_1957,Luttinger1964} the electric current density $\bm{j} =(j_x,j_y)$ is obtained from  the electric field $\bm{E} = (E_x, E_y)$ and the temperature gradient $\nabla T$ as
\begin{eqnarray}
 j_{\nu} = L_{11}^{\nu} E_\nu  - L_{12}^{\nu} \nabla_\nu T/T
                                        \; ,  \quad {\rm for}\ \nu=x, y  \; , 
   \label{eq:j}
\end{eqnarray}
where $L_{11}^{\nu}$ is the electrical conductivity $\sigma_\nu$\cite{Katayama2006_cond}  and $L_{12}^{\nu}$ is the thermoelectric conductivity in the $\nu$-direction. From  Eq.~(\ref{eq:j}), the Seebeck coefficient $S_\nu$ is given by
\begin{eqnarray}
 S_{\nu}(T) &=& \frac{L_{12}^{\nu}}{T L_{11}^{\nu}}      \; . 
   \label{eq:S}
\end{eqnarray}
We calculate $L_{11}^{\nu}$ and $L_{12}^{\nu}$ using the Sommerfeld--Bethe relation  in the same way as we performed in the case of uniaxial pressure.\cite{YS_Ogata_cond-mat} Details are shown in Appendix B.\cite{Ogata_Fukuyama}
 The quantities $L_{11}^{\nu}$ and $L_{12}^{\nu}$ are calculated  from the spectral conductivity $\sigma_{\nu}(\epsilon,T)$  defined  by  
\begin{eqnarray}
 \sigma_{\nu}(\epsilon,T) &\equiv&
 \frac{e^2 }{\pi \hbar N}  \sum_{\gamma, \gamma'}  \sum_{\bk}
  v^\nu_{\gamma \gamma'}(\bk)^* 
  v^{\nu}_{\gamma' \gamma}(\bk) \nonumber \\
    \nonumber \\
   &  \times &
     \frac{\Gamma_\g}{(\ep - E_{\gamma}(\bk))^2 + \Gamma_\g^2} \times 
 \frac{\Gamma_{\g'}}{(\ep - E_{\gamma'}(\bk))^2 +  \Gamma_{\g'}^2}
  \; ,  \nonumber 
  \label{eq:spectral_def}
\\
\end{eqnarray}
 with $v^{\nu}_{\gamma \gamma'}(\bk)  =  \sum_{\alpha \beta}$
 $d_{\alpha \gamma}(\bk)^* (\partial h_{\alpha \beta}/\partial k_{\nu})
 d_{\beta \gamma'}(\bk).$
$\Gamma_\g$ denotes the damping of electrons in the $\g$ band. $h = 2 \pi \hbar$ and $e (>0)$ are  Planck's constant  and the electric charge,  respectively.  
 
Since the spectral conductivity can be decomposed into intraband and interband 
contributions, the conductivity~\cite{Katayama2006_cond} 
 and the Seebeck coefficient are also expressed as
\begin{subequations}
\begin{eqnarray}
\sigma_\nu(T) &=& \sum_{\g,\g'} \sigma_{\nu}^{\g\g'}(T) \; ,
\label{sigma_comp}
\end{eqnarray}
\begin{eqnarray}
 S_\nu(T) &=& \sum_{\g,\g'} S_{\nu}^{\g\g'}(T) \; , 
\label{S_comp}
\end{eqnarray}
\end{subequations}
with $(\g,\g')$=(1,1), (1,2), (2,1), and (2,2), where $\g, \g'=$ 1 and 2  
correspond to the energy bands with $E_1(\bk)$ and $E_2(\bk)$, 
respectively ( {\it See} Appendix B). 

To calculate the spectral conductivity, we need the $T$ dependence of $\Gamma_\g$. As in a previous paper,\cite{YS_Ogata_cond-mat} we assume 
\begin{eqnarray}
\Gamma_{\g}  = \Gamma + \Gamma_{\rm ph}^{\g} \; ,
\label{eq:eq13a}
\end{eqnarray}
where $\Gamma$ comes from  impurity scattering 
and $\Gamma_{\rm ph}^\g$  from  phonon scattering:
\begin{subequations}
\begin{eqnarray}
  \Gamma_{\rm ph}^\g &=& C_0R \times T|\xi_{\g,\bk}|  \; , 
 \label{eq:eq16a}
        \\ 
R &=& \frac{\lambda}{ \lambda_0} \; ,  
 \label{eq:eq16b} 
\end{eqnarray}
 \end{subequations}
 where $\lambda = |g_{\bq}|^2/\omega_{\bq}$,   $\xi_{\g,\bk} = E_{\g}(\bk) - \mu$,  $C_0 = 6.25\lambda_0/(2\pi v^2)$,  and $\lambda_0/2\pi v = 0.1$.~\cite{Suzumura_PRB_2018}  $\lambda_0$ corresponds to  $\lambda$  for an organic conductor\cite{Rice,Gutfreund} and  $\lambda$ becomes  independent of $|\bq|$  for a small $|\bq|$.  $R$ is taken as a parameter.  We take $\Gamma = 0.0005$ and  $R =1$ in the present numerical calculation.~\cite{Suzumura_JPSJ2021,YS_Ogata_cond-mat}

\begin{figure}
  \centering
\includegraphics[width=6cm]{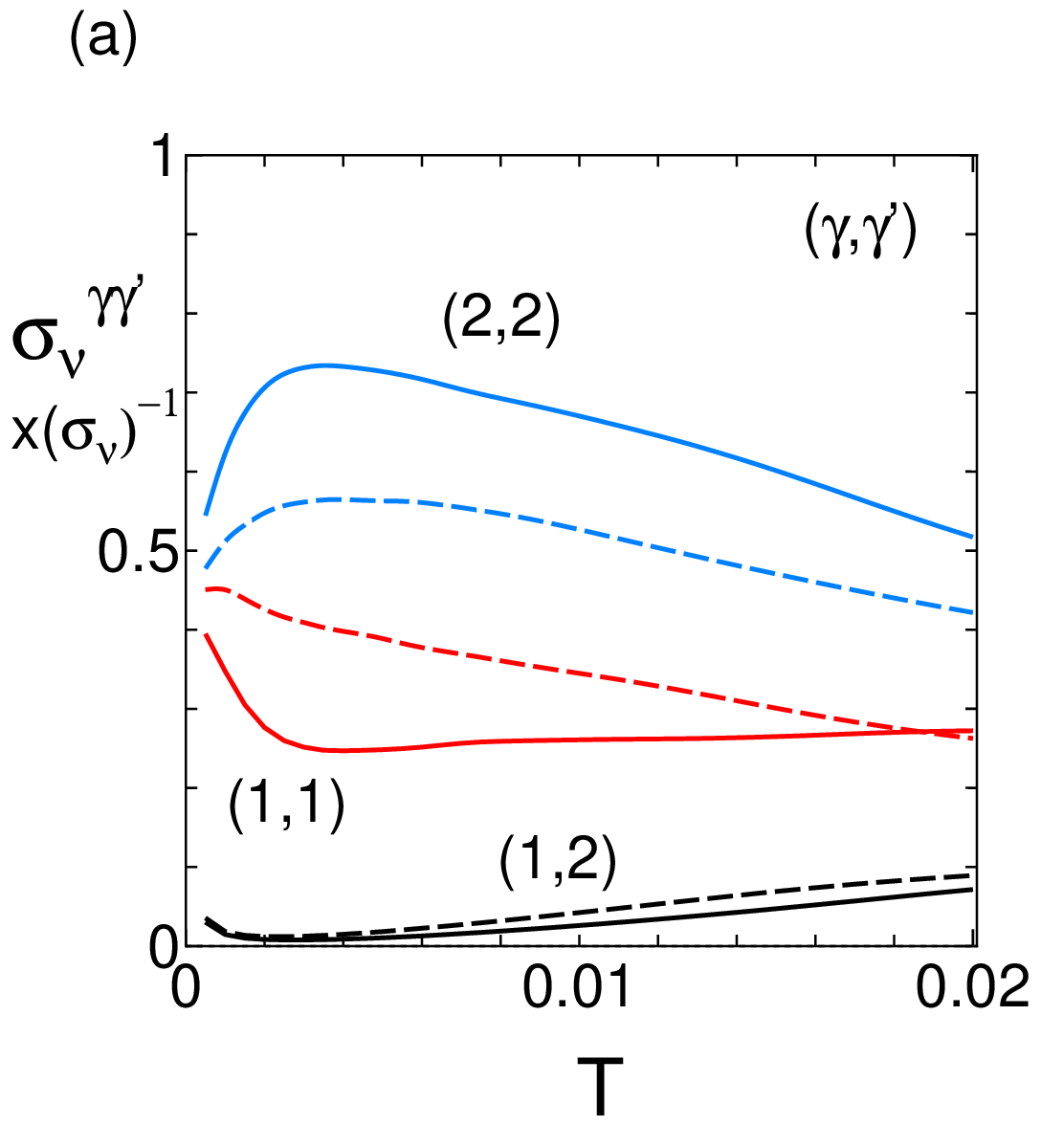}
\includegraphics[width=6cm]{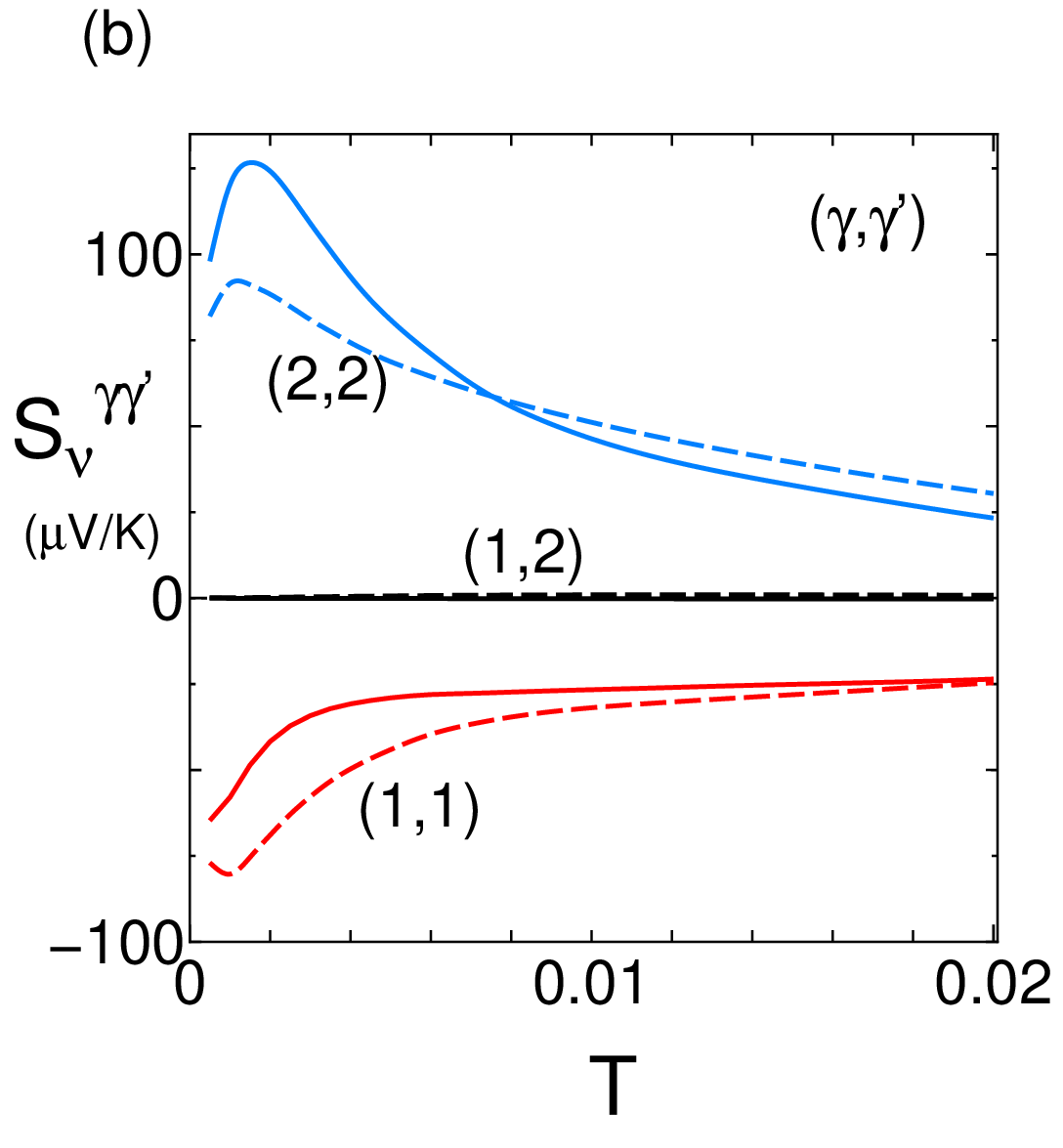}\\
     \caption{(Color online )
$T$  dependence of  the components  $(\g, \g')$ of the (a) conductivity  and 
 (b) Seebeck coefficient. The solid  and dashed lines  correspond  
 to $\nu = x$ and $\nu = y$, respectively.}
\label{fig4}
\end{figure}

Figure \ref{fig4}(a)  shows the  $T$ dependence of  $\sigma_{\nu}^{\g\g'}$. We can see that  the component of (2,2) coming from the valence  band is  larger than that of (1,1) from the conduction band, whereas  the interband contribution (1,2)  is  much smaller. This is understood   from  the  DOS of Dirac electrons, which    is inversely proportional to the velocity of the Dirac cone.  As shown in Fig.~\ref{fig3},  the DOS  of the valence band $(\omega < \muz)$  is lower than that of the conduction band $(\omega > \muz)$.  Thus,  the velocity of the valence band is higher than  that of the conduction band, leading to
 $\sigma_{\nu}^{22}(T) > \sigma_{\nu}^{11}(T)$. 
   Figure \ref{fig4}(b) shows  the $T$ dependence of  $S_{\nu}^{\g\g'}$. It is natural that the component of (1,1) is negative since it comes from the  conduction band (or electrons), whereas that of (2,2) is positive since it comes from  the valence band (or holes).  The interband contribution (1,2) is  much smaller than the others. Thus,  $S_{\nu}(T) \simeq S_{\nu}^{11}(T) +  S_{\nu}^{22}(T)$ suggests  a competition from contributions between  the conduction  and valence bands.  Note that $S_{\nu}^{22}(T) > |S_{\nu}^{11}(T)|$, which is compatible with $\sigma_\nu^{22}(T) > \sigma_\nu^{11}(T)$. 

\begin{figure}
  \centering
\includegraphics[width=7cm]{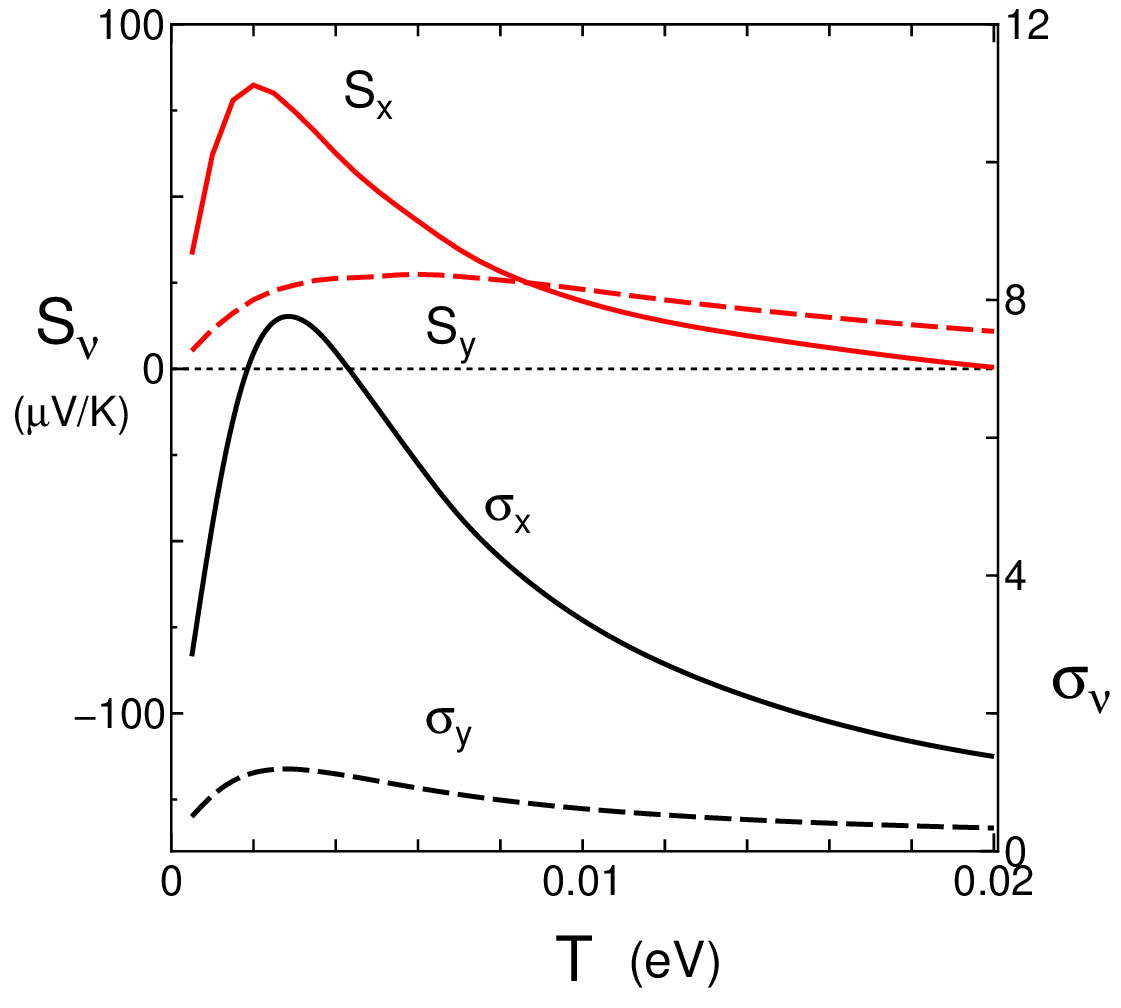}\\
     \caption{(Color online )
$T$  dependence of the Seebeck coefficients $S_x$ and $S_y$ (left axis)
 and the corresponding conductivities  
 $\sigma_x$ and $\sigma_y$ (right axis).
The unit of $\sigma_\nu$ is taken as $e^2/\hbar$.
These quantities are calculated using $\mu(T)$ shown in the inset of  Fig.~\ref{fig3}.
}
\label{fig5}
\end{figure}

 Next, we examine the $T$  dependence of the Seebeck coefficients $S_x$ and $S_y$ as shown in Fig.~\ref{fig5}. It is shown that both $S_x$ and $S_y$ are positive at finite temperatures  and take a maximum at $T = 0.002$ and $T = 0.006$, respectively.  The positive $S_\nu$  originates from $S_{\nu}^{22}(T) > |S_{\nu}^{11}(T)|$ as discussed above.
  This is also understood  from the behavior of $\mu (T)$ in the  inset of Fig.~\ref{fig3}, which shows that $\mu(T)$ is lower than $\muz$, leading to the hole-like behavior at finite temperatures. Noting that the decrease in $\mu (T) - \muz \simeq - 0.0002$  at $T=0.002$, it is suggested  that  a very small amount of  electron doping gives rise to  $S_\nu < 0$. 

 Figure \ref{fig5} also shows the corresponding conductivities,  $\sigma_x$ and $\sigma_y$. Both   $\sigma_x$ and $\sigma_y$ remain finite 
 at $T = 0$ owing  to a quantum effect~\cite{Suzumura_JPSJ_2014} and take a maximum  around  $ T  \simeq  0.003$. 
 At lower temperatures, 
the increase in $\sigma_{\nu}$ 
 originates from  the Dirac cone, which gives a linear increase  in DOS 
  measured from  the chemical potential. 
 At higher temperatures, the decrease in $\sigma_{\nu}$ 
 originates from the enhancement of the phonon scattering 
[Eq.~(\ref{eq:eq16a})].
The relation  $\sigma_x > \sigma_y$ comes  from the anisotropy of the Dirac cone,\cite{Suzumura_JPSJ_2014} which gives   $v_x > v_y$, as shown in Fig.~\ref{fig2}(e). There is a similarity of the $T$ dependence between $S_\nu$ and $\sigma_{\nu}$  since both quantities depend on  the spectral conductivity. However,  there is a difference in origin between  $S_\nu$ and $\sigma_{\nu}$.  $S_\nu$ is determined by the effects of competition  between the valence and conduction bands, whereas  $\sigma_{\nu}$ is obtained from the  additive effects of the valence and conduction  bands. 

\begin{figure}
  \centering
\includegraphics[width=7cm]{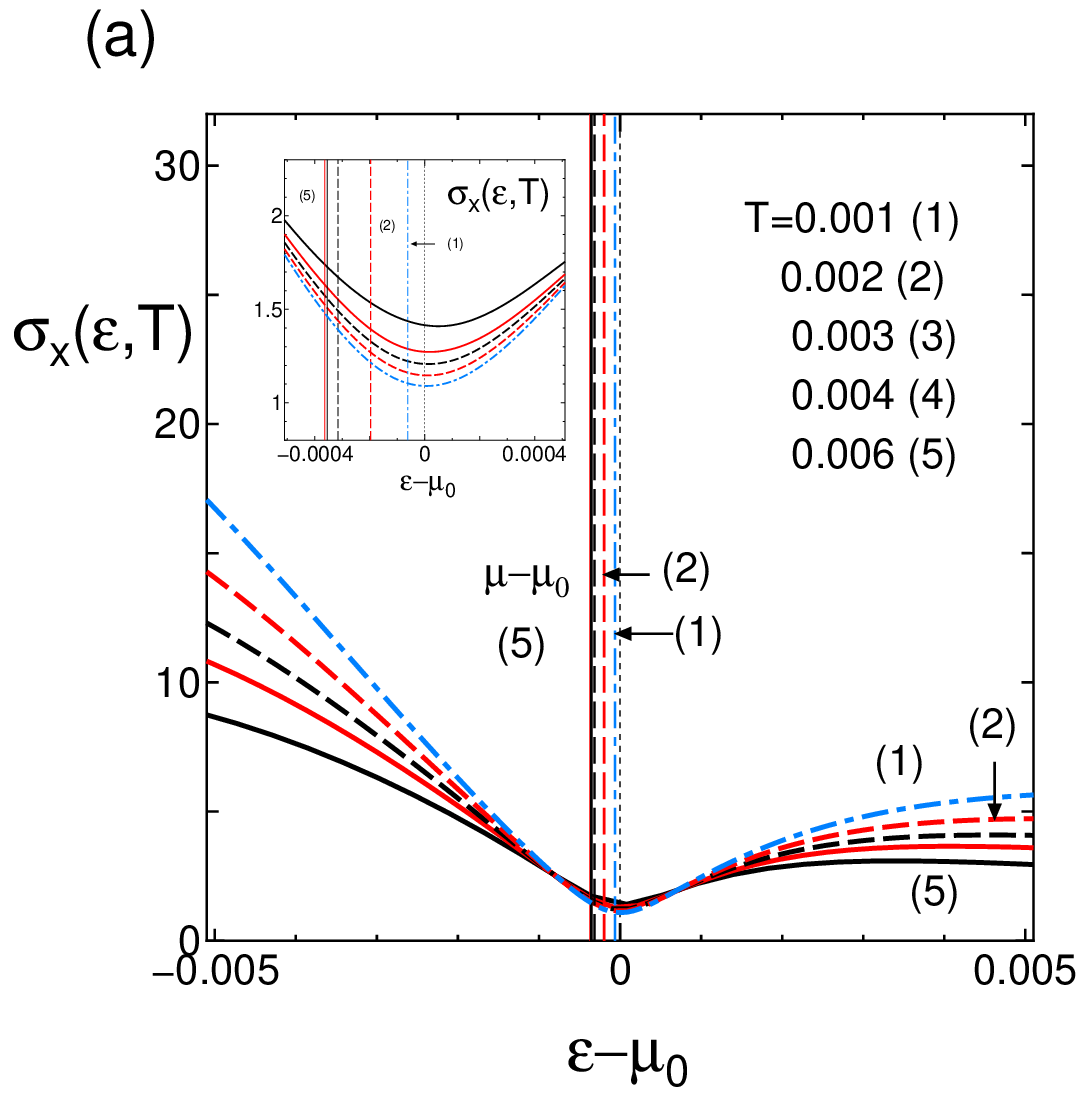}\\
\includegraphics[width=7cm]{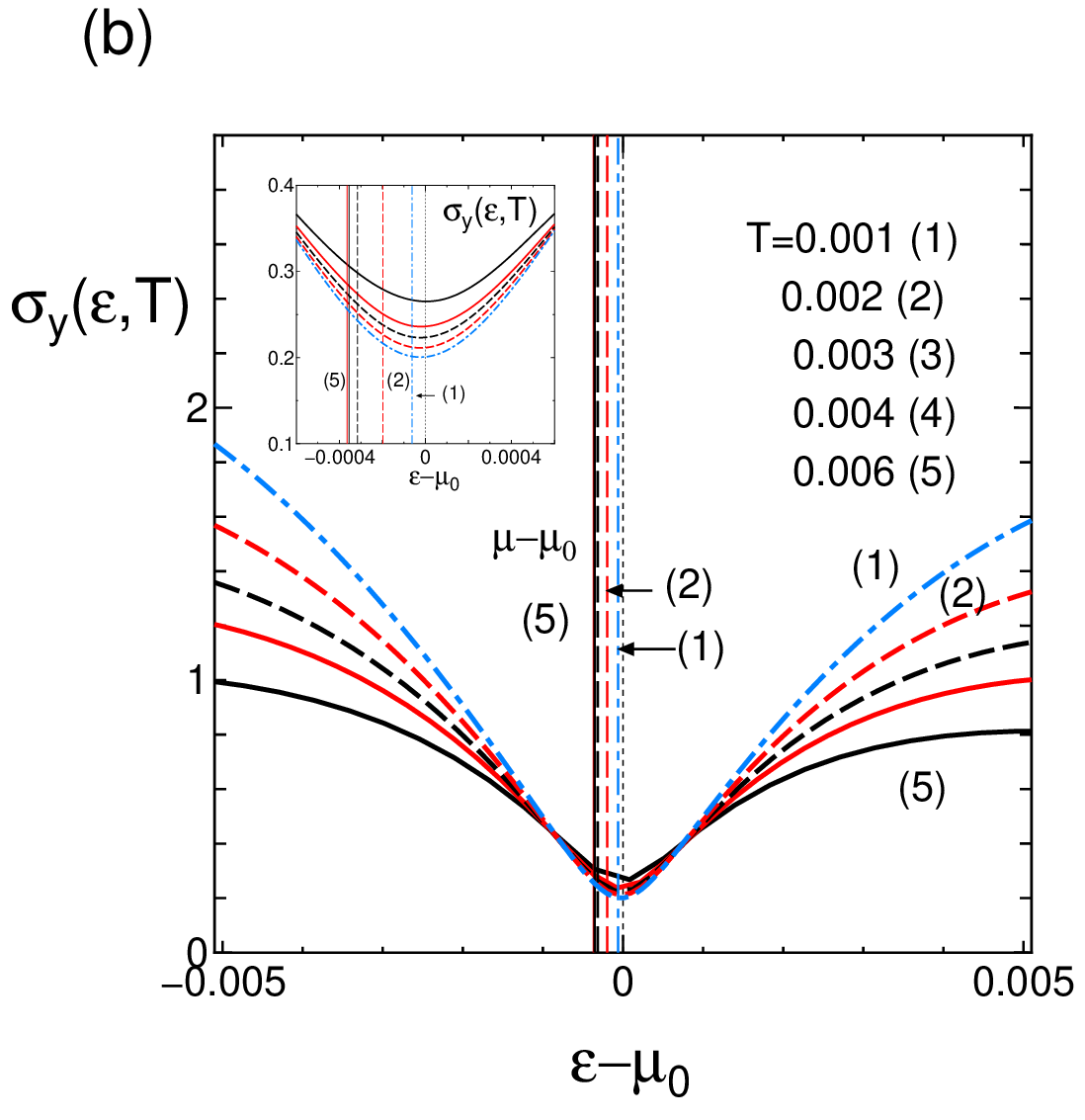}\\
     \caption{(Color online)
$T$ dependence of the spectral conductivities \spcx (a) and \spcy (b)  as a function of $\ep - \muz$ with  fixed temperatures $T =$ 0.001 (1), 0.002 (2), 0.003 (3), 0.004 (4) and 0.006 (5). The vertical line shows $\mu - \muz$ for the corresponding temperatures. The inset  denotes the respective $\sigma_\nu(\ep,T)$ with a magnified scale. }
\label{fig6}
\end{figure}

\subsection{Spectral conductivity }
Here, let us examine the sign of $S_\nu$  using the spectral conductivity $\sigma_{\nu}(\epsilon,T)$. When we expand $\sigma_{\nu}(\epsilon,T)$ as
\begin{eqnarray}
  \sigma_\nu(\ep,T) &=&  \sigma_\nu(\mu,T)
       + \sigma'_{\nu}(\mu,T)(\ep - \mu)  
                    \nonumber    \\
    && + \frac{1}{2} \sigma''_{\nu}(\mu, T)  (\ep - \mu)^2    + \cdots  ,
      \label{eq:L11_T} 
\end{eqnarray}
the thermoelectric conductivity $L_{12}^{\nu}(T) $ is given by~\cite{YS_Ogata_cond-mat}           
\begin{eqnarray}
  e L_{12}^{\nu}(T) &=&  
       - \frac{\pi^2}{3} \sigma'_{\nu}(\mu,T)T^2
       - \frac{7\pi^4}{90} \sigma'''_{\nu}(\mu,T)T^4 
                  \nonumber         \\
        &&    + \cdots . 
      \label{eq:L12_T}         
\end{eqnarray}
Note that the first term corresponds to the Mott formula when $T\rightarrow 0$,~\cite{Mahan1980}  and it indicates that  $S_\nu(T) < 0$ ($S_\nu(T) > 0$)  when   $\sigma'_{\nu}(\mu,T) > 0$   ($\sigma'_{\nu}(\mu,T) < 0$).

Figures \ref{fig6}(a)  and  \ref{fig6}(b) show  the spectral conductivities \spcx   and \spcym, resprctively   as a function of $\ep - \muz$ at several temperatures.
 They show a common feature that the minimum of \spcn ($\nu = x$ and $y$) is found  at $\ep - \muz \simeq 0$. The effect of temperature is seen for  $|\ep - \muz| > 0.001$. There is a small deviation of the minimum of  \spcn from $\muz$.
 It can be shown analytically that  \spcn  takes the minimum at $\ep = \muz$  in the two-band model with the linear dispersion regardless of the tilting. 
   Thus, the deviation of the minimum of \spcx from $\ep = \muz$ is ascribed to the  energy band around the Dirac point, which  deviates  from the linear dispersion. The vertical lines  denote  $\ep=\mu(T)$ 
 for the respective  temperatures, which reads  the tangent of \spcn at $\ep=\mu(T)$.  Since $\mu(T) < \muz$,
 $\sigma'_\nu(\ep, T) < 0$  at  $\ep = \mu(T)$,  suggesting $S_\nu(T) > 0$ at any temperature.
  We can see that \spcx $>$ \spcym, which is consistent with $v_x>v_y$ as discussed above. Furthermore, we find that the anisotropy of \spcx  with respect to 
 $\ep - \muz$ is larger than that of \spcym.  This can be understood from the energy bands shown in  Figs.~\ref{fig2}(c) and \ref{fig2}(d), where anisotropy along $\delta k_x$ is much larger than that along $\delta k_y $  owing  to the tilting of the Dirac cone along the $k_x$ direction. 
  Finally, the magnitude of \spcn for $\ep-\muz<0$ is larger than \spcn for $\ep-\muz>0$, which is consistent with the asymmetry of DOS as discussed in Fig.~\ref{fig4}(a).

\section{Summary and Discussion}
We have examined the $T$ dependence of the Seebeck coefficient for the Dirac electrons in \ET \; under hydrostatic pressure  using the improved TB model obtained  from the first-principles DFT calculations with the experimentally determined  crystal structure. 
   To the best of our knowledge, the  TB model presented in this  paper is unique for explaining the Seebeck coefficient of the present organic conductor.  Main results of this study are as follows:
   (i) The Seebeck coefficient is positive 
 in both $x$- and $y$-directions, i.e., $S_x>0$  and $S_y>0$, indicating the hole-like behavior, which is consistent with $\mu(T) - \muz < 0$. 
   However, note that the sign of the Seebeck coefficient is determined not only from the sign  of $\mu(T) - \muz$  but also from the difference between the energy dispersions of the conduction and valence bands. Although the quantitative understanding of such competition between the conduction and  valence bands is complicated, it is demonstrated that  an effective  method of comprehending the sign of the Seebeck coefficient is to examine the energy derivative of  the spectral conductivity, $\sigma_\nu(\ep,T)$, at $\ep = \mu(T)$. 
   (ii) Another result obtained from  the transfer energies of the present model is  $\sigma_x > \sigma_y$ as seen from the behavior of the Dirac cone [Fig.~\ref{fig2}(e)]. This is in sharp contrast to that of  the model of the extended H\"uckel method, which shows  $\sigma_y > \sigma_x$.~\cite{Suzumura_JPSJ2021}
   (iii)  The analysis of the spectral conductivity suggests that $S_\nu < 0$ can be realized by  a small amount of  electron doping, i.e., 
 $\mu - \muz \simeq 0.0001$.  

   Let us compare our calculation  result of $S_\nu$  with the experimental 
 results of \ET\ under hydrostatic pressures. So far,  two experiments have been reported.~\cite{Konoike2013,Tajima_Seebeck}
   One is a measurement at $P$ = 1.5 GPa,~\cite{Konoike2013} which shows that the positive $S_\nu$  has a small peak at around 50 K and then decreases as the temperature decreases up to 2 K. No sign change at low temperatures is predicted. 
 This is consistent with our theoretical results of $S_x>0$ and $S_y>0$. 
Moreover, the existence of a maximum of $S_\nu$ is also consistent with our results. 
   Another experiment is a measurement performed at $P$ = 1.9 GPa, where  a sign change of $S_x$ occurs at around 3 K and the positive $S_x$ shows a peak at high temperatures. Such a sign change depends on the choice of samples,  and we speculate that this sign change can be understood by assuming a small amount of electron doping, since the variation of the chemical potential  is $\sim 0.0001$ for $T = 0.001$ as discussed in this paper.  Thus, the present TB model obtained from the DFT calculations explains  successfully the Seebeck coefficient obtained from the experiments  of \ET, 
which depend on the doping amount.
   Finally, we comment on the anisotropy of the conductivities shown in Fig.~\ref{fig5}, where $\sigma_x$ is much larger that $\sigma_y$.
  The experimental results show that $\sigma_x$ is about twice as much as $\sigma_y$  and their ratio is expected to increase for a clean sample.\cite{Tajima_conductivity}
 
\acknowledgements
We thank N. Tajima for helpful discussions and sending us the data on the Seebeck coefficient  of \ET \; at 1.9 GPa. This work was supported by  Grants-in-Aid for Scientific Research (Grants No. JP23H01118 and JP23K03274), 
JST CREST Grant No. JPMJCR18I2, and JST-Mirai Program Grant Number JPMJMI 19A1.

\appendix
\section{Matrix elements  in the TB model}
The TB model $H_0$ in Eq.~(\ref{eq:Hij})  is expressed as 
\begin{eqnarray}
H_0 &=& \sum_{\bk}  \sum_{\alpha, \beta}
 h_{\alpha, \beta}(\bk)  a^{\dagger}_{\alpha}(\bk) a_{\beta}(\bk) 
\; .
\label{eq:HijApp}
\end{eqnarray}
 The Fourier transform for the operator $a_{j,\alpha}$ is defined as  $a_{j,\alpha} = 1/N^{1/2} \sum_{\bk} a_{\alpha}(\bk) \exp[ i \bk \cdot \bm{r}_j]$, where $\bk = (k_x,k_y)$ for the  2D case  with  $ - \pi < k_x, k_y \leq \pi$  and the lattice constant is taken as unity. Using a basis of four molecules in the unit cell ($\alpha$ = 1, 2, 3, and 4), we  write  the matrix element $h_{\alpha,\beta}(\bk)$ as  
\begin{subequations}
\begin{eqnarray}
h_{12}(\bk) &=& a_3 + a_2 Y\;, \\
h_{13}(\bk) &=& b_3 + b_2 X\;, \\
h_{14}(\bk) &=& b_4 Y + b_1 XY\;,\\ 
h_{23}(\bk) &=& b_2 + b_3 X\;, \\
h_{24}(\bk) &=& b_1 + b_4 X\;, \\
h_{34}(\bk) &=& a_1 + a_1 Y\;, \\
h_{11}(\bk) &=& t_{22}(\bk) = a_{1d} (Y + \bar{Y})\;, \\
h_{33}(\bk) &=&  a_{3d} (Y + \bar{Y})+ V_{\rm B}\;, \\
h_{44}(\bk) &= &a_{4d} (Y + \bar{Y})+ V_{\rm C}\;, 
\end{eqnarray}
\end{subequations}
and $h_{\alpha,\beta}(\bk) = h_{\beta,\alpha}^*(\bk)$,  where   
$X=\exp[i kx] = \bar{X}^*$  and  $Y= \exp[i ky] = \bar{Y}^*$.  $k_x$ corresponds to the direction perpendicular to the molecular stacking axis.

\section{Spectral conductivity and  Seebeck coefficient}
The electrical  and   thermoelectric conductivities are 
 respectively written as 
\begin{eqnarray}
L_{11}^{\nu} &=&  \sigma_{\nu}(T) = 
              \int_{- \infty}^{\infty} d \ep  
            \left( - \frac{\partial f(\ep) }{\partial \ep} \right)
         \times  
\sigma_{\nu}(\epsilon,T)   \;, 
                             \nonumber          \\
   \label{eq:L11}
\end{eqnarray}
\begin{eqnarray}
 L_{12} &=&  \frac{-1}{e}
              \int_{- \infty}^{\infty} d \ep  
  \left( - \frac{\partial f(\ep) }{\partial \ep} \right)
         \times (\ep - \mu) \sigma_{\nu}(\ep,T)  \; .
        \nonumber \\
   \label{eq:L12}
\end{eqnarray}
The spectral conductivity $\sigma_{\nu}(\epsilon,T)$  with $\nu = x$ and  $y$ is calculated as~\cite{YS_Ogata_cond-mat} 
\begin{eqnarray}
\sigma_{\nu}(\epsilon,T) &=&   
      \sum_{\gamma, \gamma'}\sigma_{\nu}^{\gamma \gamma'}(\epsilon,T)
                     \; ,          
  \label{eq:spectral}
\end{eqnarray}
where 
\begin{eqnarray}
 \sigma_{\nu}^{\gamma \gamma'}(\epsilon,T) &=&
 \frac{e^2 }{\pi \hbar N} 
  \sum_{\bk}
  v^\nu_{\gamma \gamma'}(\bk)^* 
  v^{\nu}_{\gamma' \gamma}(\bk) \nonumber \\
    \nonumber \\
   &  \times &
     \frac{\Gamma_\g}{(\ep - E_{\gamma}(\bk))^2 + \Gamma_\g^2} \times 
 \frac{\Gamma_{\g'}}{(\ep - E_{\gamma'}(\bk))^2 +  \Gamma_{\g'}^2}
  \; .  \nonumber 
  \label{eq:spectral_comp}
\\
\end{eqnarray}
$v^{\nu}_{\gamma \gamma'}(\bk)$ denotes  a  matrix element of the velocity  
 given by $v^{\nu}_{\gamma \gamma'}(\bk)  =  \sum_{\alpha \beta}$
 $d_{\alpha \gamma}(\bk)^*  (\partial h_{\alpha \beta}/\partial k_{\nu}) d_{\beta \gamma'}(\bk).$ 
   Using $\sigma_{\nu}^{\gamma \gamma'}(\epsilon,T)$, 
 we obtain the Seebeck coefficient $S_\nu$    by 
\begin{eqnarray}
S_{\nu}(T) = \frac{L_{12}^\nu}{T L_{11}^{\nu}}
    =\sum_{\g, \g'} S_{\nu}^{\g\g'}(T)
     \label{eq:S}
\end{eqnarray}
where
\begin{eqnarray}
 S_{\nu}^{\g\g'}(T) &=& \frac{1}{T L_{11}^{\nu}} \times
                                     \nonumber \\
   &&  \frac{-1}{e}
              \int_{- \infty}^{\infty} d \ep  
  \left( - \frac{\partial f(\ep) }{\partial \ep} \right)
         \times (\ep - \mu) \sigma_{\nu}^{\g\g'}(\ep,T)
                    \; ,       \nonumber \\
   \label{eq:S_comp}
\end{eqnarray}



\end{document}